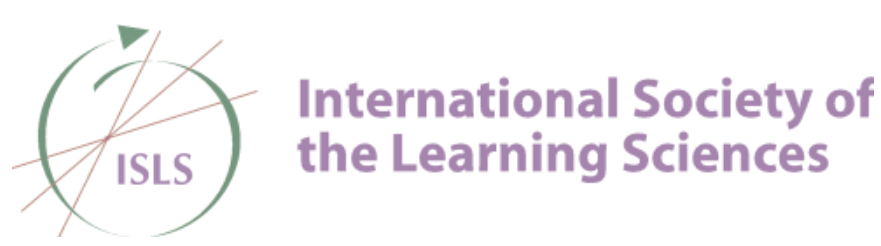

# What If AI Carried Her Imagination? Black Girls as Creators in an AI Storytelling Weekend Program

Chun Li, University of Pittsburgh, chl500@pitt.edu
Lauren Brown, Manchester Youth Development Center, lbrown@mydc.org
Hubert Asare, University of Pittsburgh, hoa30@pitt.edu
Shawna Patterson, Manchester Youth Development Center, spatterson@mydc.org
Dennis Henderson, Manchester Youth Development Center, dhenderson@mydc.org
Ericka Roland, University of Texas at Arlington, ericka.roland@uta.edu
tara Nkrumah, Arizona State University, tnkrumah@asu.edu
Angela E.B. Stewart, University of Pittsburgh, angelas@pitt.edu

**Abstract:** This paper presents the design and outcomes of a seven-weekend AI storytelling program developed for Black girls aged 10-12. Grounded in Afrofuturism and Black feminist thought, the program adopted AI-enabled counter-storytelling, supported the development of foundational AI literacies, and fostered future-oriented imagination. Activities included brainstorming AI-related topics, developing character and story plots, and delivering collaborative group presentations. Drawing on the analysis of learners' artifacts from the case study, findings show that participants created Afrofuturist narratives rooted in their identities and everyday experiences. At the same time, they developed core AI literacies, including prompt engineering, bias critique, and awareness of data privacy. This program demonstrates that integrating Afrofuturist storytelling with generative AI in informal learning spaces can be a powerful approach for engaging Black girls in computer science education.

## Introduction

What if a Black girl created a robot that blocked the flooding in her neighborhood? What if her streets were forever clean of the refuse from the river? What if Black girls led the field of technology?

These "what if" style questions were the core of our Artificial Intelligence (AI) storytelling weekend program, where ten Black girls created speculative stories and learned about AI. Storytelling is a powerful tool in computer science (CS) education, offering opportunities for learners' social and cognitive development through creativity, critical thinking, and collaboration (Cabrera et al., 2025). As AI becomes increasingly integral to CS education, storytelling pedagogies offer a meaningful pathway into AI literacy (Ng et al., 2022). They help learners grasp core concepts such as prompt engineering, while also encouraging critical engagement with AI's social implications. Through this process, storytelling nurtures critical literacies, enabling learners to not only understand AI but to question and reshape it through their own narratives (Ellis et al., 2018; Scott et al., 2015).

Within the United States context, Black girls remain one of the most underrepresented groups in the computing field (Li et al., 2023). Historic and systemic racism and patriarchy have contributed to the marginalization of Black girls in CS and society more broadly (Ashcraft et al., 2017). In contrast, Afrofuturistic pedagogy, a lens centering Black experiences and envisioning a technological future, offers powerful inspiration for valuing Black girls' identities as assets in their learning (Ellis et al., 2018). Guided by this lens and the provocation "What if the field of technology were led by Black women and girls?" our program challenged dominant narratives in computing by positioning Black girls as creators of future technologies, inviting them to use AI tools for self-expression, and supporting them in developing AI literacy.

In this work, we advocate for the incorporation of Afrofuturism as a theoretical framing and storytelling as pedagogical tools, while using AI tools as a vehicle for advancing Black girls' engagement in technology. Our study was guided by the research question: How can Afrofuturism, as a theoretical lens, inform AI-enabled storytelling practices with Black girls? To answer this question, we employ a case-study approach, examining learners' artifacts to capture the interplay between program and learning outcomes. Within the context of learning science, this work makes two key contributions. First, it introduces a CS pedagogical approach that is built on Afrofuturist and Black feminist thought, presenting a curriculum that blends narrative-building with AI. Second, it centers Black girls as creators of speculative futures using AI tools, highlighting their creativity and identity expression, and challenges traditional norms that prioritize technical competency alone (Scott et al., 2015).

## Conceptual framework

In this section, we review related work in Afrofuturism and Black feminist thought, focusing on the application of counter-storytelling in CS education.

## Afrofuturism, Black feminist thought, storytelling, and counter-storytelling

Afrofuturism intersects with imagination, technology, and liberation, offering a way of envisioning possible futures through a Black and Afrocentric lens. As a cultural movement, Afrofuturism is a creative mode for responding to the historical trauma brought by racism and colonialism (Strong et al., 2023). For example, the movie *Black Panther* presents an Afrofuturist vision of the fictional country Wakanda, posing the question of Black liberation: "What if Africa had never been colonized?" (Dando et al., 2019). In this movie, narratives of Black people are reconstructed to center connections to African ancestors and reimagine a healing future (Ellis et al., 2018). In Afrofuturist production, technology is central to imagining liberatory futures through technoculture and science fiction (Strong et al., 2023). As a pedagogical approach, Afrofuturism in CS education emphasizes Black learners' vision of themselves in the future (Jones & Howard, 2022).

Complementing this, Black feminist thought emphasizes that Black experiences are vital sources of knowledge in understanding systems of oppression and in envisioning paths toward liberation (Collins, 2022). Thus, the synthesis of Afrofuturism and Black feminist thought is considered Afrofuturist feminism, using technology as a lens to imagine liberatory futures for Black women and girls (Morris, 2012).

Afrofuturism can be easily fused with storytelling (Mcgee & White, 2012). Historically, Black people have long created stories that express themselves, transfer knowledge, and construct dreams of freedom (Kaler-Jones, 2022). As the Afrofuturist novelist, Octavia Butler, said, "Every story I create, creates me. I write to create myself." Storytelling is a fundamental aspect of collective knowledge, serving as a primary means for sharing knowledge across generations (Alsaleh & Vasanth, 2025). However, not everyone has had equal opportunities to be heard. For those at the margins of society and history, counter-storytelling, a critical race methodology, offers a way to challenge dominant narratives and "counter" deficit-based representations. Counter-storytelling centers marginalized knowledge and disrupts majoritarian narratives while affirming traditions of social, political, and cultural survival and resistance (Solórzano & Yosso, 2002). For Black girls, counter-storytelling offers an opportunity to name the oppression and speak back to it (Kaler-Jones, 2022), since conscious acts and expressions of joy are vital forms of resistance against racism and violence (Lu and Steele, 2019). Scholars recognized that counter-storytelling is a subversive tool of resistance and a construct of Black joy (Adams, 2022). In addition to these justice-oriented concerns, prior research further confirms the power of narrative to spark the interests of learners in CS and disciplinary identification (Pinkard et al., 2017).

In this work's landscape, Afrofuturism presents an optimistic act of creation, envisioning futures shaped by technology, and the counter-storytelling approach encourages Black girls to create counter-narratives of traditional characters. Together, these lenses informed the program design.

## Black girls as AI creators: the storytelling program

In this section, we describe the design and implementation of the program, Black Girls as AI Creators, where learners engaged in activities that were scaffolded to develop creativity, agency, and AI literacy. Researchers participated as both observers and facilitators, shaping how we interpreted participants' creative practices.

### MYDC: history and values

Based on a research-practice partnership (RPP) between universities and a local community education center (Coburn & Penuel, 2016), this program took place from January to March 2025 in the Manchester Youth Development Center (MYDC), located in Pittsburgh, Pennsylvania, United States. MYDC is a nonprofit, community-based youth development and education organization that has primarily served Black communities over the past five decades. Media creation is a significant emphasis in MYDC's programming, focusing on creating artifacts for self-expression and identity development. This program was conceptualized in RPP between MYDC and a research team of interdisciplinary scholars from fields such as Information Science, Learning Science, and Human-Computer Interaction. Together, we co-designed and implemented the program's curriculum to align with community goals for culturally responsive and empowering technology learning experiences.

### Program design

The program was facilitated by two educators from MYDC, both Black American women; two graduate students from a local university: one an Asian woman and the other an African man. The research team was led by three Black American women faculty members who developed the theoretical and pedagogical foundations of this work. The RPP brought together MYDC and universities, enabling iterative refinement of the curriculum while also supporting the professional learning of the participating educators. The curriculum was structured across four modules that combined AI foundational concepts and privacy discussions, open-ended AI exploration (e.g., using AI tools to research Black historical and contemporary figures), speculative "What if?" prompts development and storytelling practices, and character and story development. After seven weekends, the program concluded with a family day, during which learners shared and celebrated their creations with their families.

## Findings: Case study – learner-created artifacts

We conceptualize the analysis of learners' artifacts as a case study that illustrates how Black girls engaged with AI (Yin, 2009). Informed by Afrofuturism and counter-storytelling, we examine how learners constructed identity-centered narratives and engaged with AI, as well as their emerging AI literacy practices, particularly through their spontaneous prompt engineering. Data sources include lesson plans, facilitators' field notes, learners' AI interaction logs and oral reflections, and learner-created artifacts (e.g., generated images and narratives) collected during program sessions.

### Narrative building with and for Black Girls in AI spaces

Throughout the seven weekends, learners engaged in sustained narrative building, drawing on lived experiences and cultural references. In their brainstorming discussions, they drew inspiration from their homes, schools, and communities. This process involves transforming real-world contexts into creative narratives. Even when the stories became speculative or fantastical, they were still rooted in what the learners knew and cared about. Frequently referenced elements included everyday objects (e.g., Stanley water bottles, shoes, books), environments (e.g., campsites, schools, gyms, grasslands), and technologies (e.g., TVs, robots, phones). By centering these familiar elements, learners positioned their everyday lives as worthy of storytelling.

Black girls consistently emerged as the heroes of the stories, which is a powerful contrast to their underrepresentation as central characters in mainstream K-12 books and media (Zhou et al., 2022). When prompting for their characters, learners often used descriptions that mirrored aspects of themselves or imagined versions of themselves, such as "a rockstar fourteen-year-old girl," "an African rich girl wearing Jordan shoes," "a brown-skinned girl with a bit of freckles," and "a girl alien with boho braids." Learners constructed Black girl-centered narratives, naturally and collaboratively answering the question, "What if Black women and girls led technology?" using the creative possibilities of generative AI. Their character prompts reflected detailed self-representation and imagination, particularly in centering Black girls with specificity. The characters' building represents narrative building through visual, linguistic, and cultural cues.

The everyday experiences that learners have transferred into imaginative narratives might otherwise be challenging to produce using traditional tools. As children's ideas and imagination often outpace their fine motor skills or writing abilities, many children, especially those in late elementary or middle school, do feel discouraged when they struggle to create something "fancy", particularly in visual or narrative tasks (Golomb, 2004). Generative AI tools can bridge the gap between skills and imagination, enabling the creation of more complex characters, visuals, and storylines for youth. Guided by Afrofuturism, we attend to how these stories position Black girl characters as heroes, leaders, and innovators within futuristic and fantasy settings, reflecting rich forms of liberatory imagination. Through this work, Black girls actively constructed Afrofuturist narratives.

### Cultivating AI literacy through navigating tool limitations

Learners engaged in ongoing interactions with AI tools to generate, refine, and critique their stories and characters. Below, we present some examples and one vignette that illustrates how learners navigate the power and limitations of generative AI. To protect the privacy and dignity of minor participants, we do not include any AI-generated images created directly by the learners. Instead, we present recreated images generated by ChatGPT-4, based on the original prompts used in the program.

Seven learners chose girl protagonists like themselves in age and engaged in extensive prompt refinement, focusing especially on visual appearance. Many explicitly requested darker skin tones and authentic hairstyles that reflected their identities. Initially, learners encountered confusion when outputs did not match their intentions. For example, one learner requested a "girl with curly hair and brown eyes in pajamas," but after specifying "shorter hair" and "darker skin" regarding the first output, the system generated an image of a boy instead. Though surprised, she continued refining her prompt. Like her, many learners engaged in spontaneous prompt engineering until the outputs aligned with their expectations.

One learner wanted to create a story about "a rockstar 14-year-old girl who is a rebel and wants to make the world normal again." For the character illustration, she initially prompted: "light skin" (a common term within Black communities referring to lighter-complexioned Black individuals), "wavy brown hair, and rockstar clothes," then received an image of a white girl character, as shown in Figure 1 (a). She expected a skin tone color like hers, so she revised the prompt, asking for "a light chocolate skin." This adjustment yielded the image on the right (see Figure 1(b)). Yet the result still did not align with her vision, so she requested a full-body image by adding "full body please", expecting to receive a portrait that looked somewhat different from the last one. However, the new output reverted to a white girl, resembling the original image (see Figure 1(c)). She then, again, clarified her prompt with "an African girl," which generated the final image shown on the right (see Figure 1(d)). Satisfied with this version, she moved on to other aspects of her story design. Through the editing process, learners asserted aesthetic agency and celebrated aspirational self-representation.

**Figure 1**
*AI-generated pictures of "a rockstar girl"*

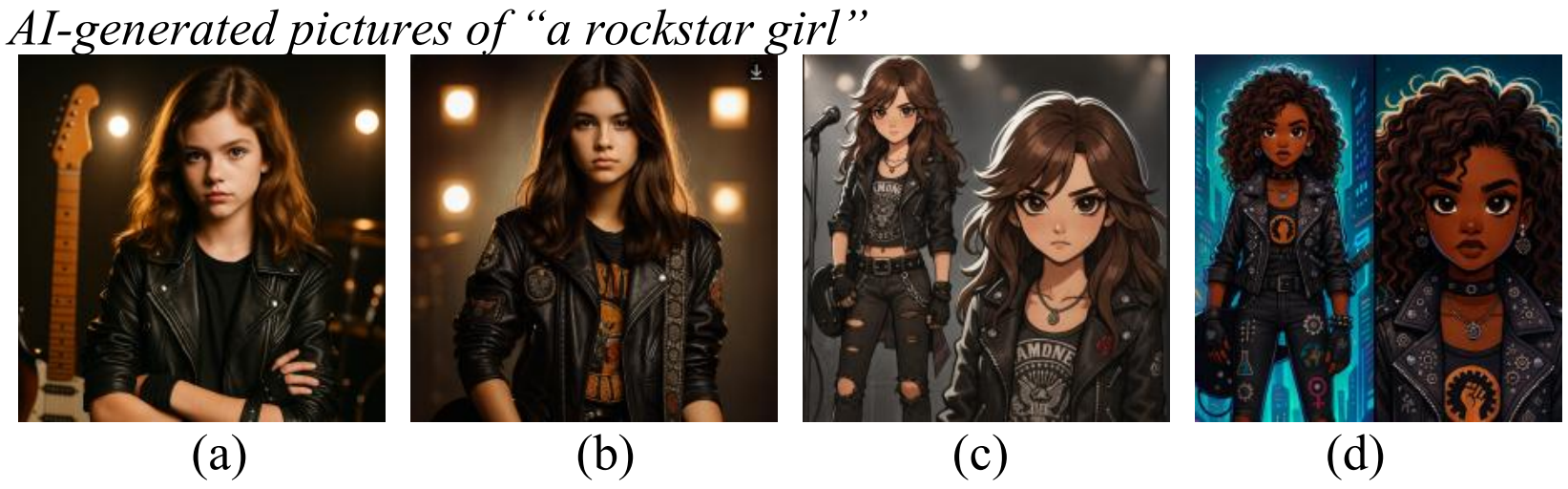

(a) (b) (c) (d)

This moment reflects learners' growing understanding of prompting and the limitations of generative AI. Nearly all who created human characters encountered implicit bias, with default outputs often depicting white or light-skinned figures. When faced with this prompting challenge, learners responded with comments like, "Oh, that's not what I wanted," and revised their prompts, most frequently using instructions like "make her darker." Learners treated the chatbot as a conversational partner, initially assuming it understood their cultural context. When it did not, they adapted their language, shifting from familiar terms to more explicit descriptors. In one case, a learner who wanted to design a Black girl figure requested "darker skin" more than six times without success, eventually adapting her phrasing to "make her look like an African American" or even "Blacker." Once satisfied with the skin tone, she moved on to refining the character's hairstyle. Such linguistic negotiation highlights both their growing awareness of AI's limitations and the need to negotiate meaning on different terms. In these moments, Black girls recognize AI as a "double-edged sword": a tool that could empower them to create outstanding artistic artifacts, yet one that could also marginalize their identities. During the family day, when discussing her story with her parents and a researcher, one learner told her parents that she had to continually refine the prompt to get the proper skin tone. Learners wanted to center themselves in a visualized speculative future while also learning to think critically about technologies.

## Discussion and conclusion

To answer our research question, how can Afrofuturism inform AI-enabled storytelling practices with Black girls? This program employed Afrofuturism as a theoretical lens to expand how Black girls interacted with AI by foregrounding their lived experience, speculative imagination, and critical resistance to dominant narratives. Through this lens, learners engaged in AI-enabled counter-storytelling while developing AI literacies. Specifically, Black girls engaged in activities to understand how AI works, to create Black girl-centered stories, to evaluate the quality of their prompts and AI-generated outcomes, and to explore AI ethics issues. Given the autonomy to create and reflect on stories about themselves, Black girls became thoughtful interpreters and producers rather than passive consumers of technology. As AI becomes an increasingly important part of youth's future, Black girls will both benefit from imagining their own technological possibilities and confronting AI's existing limitations. Early experiences like this program serve as preparation before they engage more deeply in technology in their future study and careers. Thus, learners demonstrated emerging AI literacy while also recognizing the evolving nature of generative AI.

We also encountered several challenges. Technical delays occurred when multiple learners shared a single ChatGPT account, resulting in slower response times and interruptions to the creative flow. Additionally, a few participants experienced uncertainty due to limited typing skills or unfamiliarity with digital devices, highlighting the need for more scaffolded technical support for younger learners. These challenges offer critical insights for future curriculum design and facilitation strategies in ensuring equitable engagement.

Ultimately, this Afrofuturist AI storytelling program illuminated how Black girls can position themselves as creators of imaginative futures, engaging technology with both critical insight and joy. We hope this program offers valuable perspectives for educators and curriculum designers as they respond to AI's growing influence on K-12 CS classroom dynamics, particularly by fostering learners' agency, creativity, and identities.

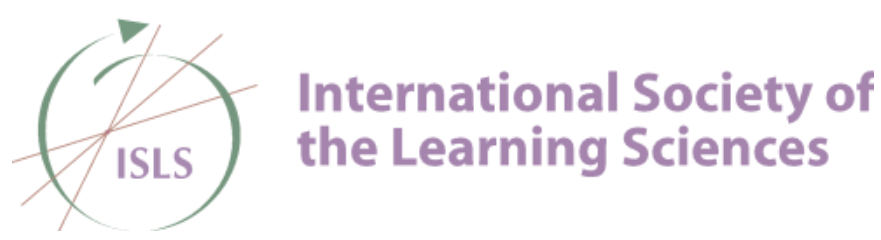